\documentclass[conference]{IEEEtran}
\IEEEoverridecommandlockouts
\usepackage{cite}
\usepackage{comment}
\usepackage{amsmath,amssymb,amsfonts}
\usepackage{algorithmic}
\usepackage{amssymb}
\usepackage{graphicx}
\usepackage{textcomp}
\usepackage{amsmath}
\usepackage{multirow}
\usepackage{xcolor}
\def\BibTeX{{\rm B\kern-.05em{\sc i\kern-.025em b}\kern-.08em
    T\kern-.1667em\lower.7ex\hbox{E}\kern-.125emX}}
\usepackage[letterpaper, left=0.75in, right=0.75in, bottom=0.75in, top=0.75in]{geometry}
\begin{document}

\title{Attention Based Feature Fusion For Multi-Agent Collaborative Perception\\}

\author{\IEEEauthorblockN{Ahmed N. Ahmed\IEEEauthorrefmark{1},
Siegfried Mercelis\IEEEauthorrefmark{1}, Ali Anwar\IEEEauthorrefmark{1}}

\IEEEauthorblockA{\IEEEauthorrefmark{1}Department of Electronics and ICT Engineering, IDLab - University of Antwerp - imec, Antwerp, Belgium}
\IEEEauthorblockA{ahmed.ahmed@uantwerpen.be}}

\maketitle

\begin{abstract}
In the domain of intelligent transportation systems (ITS), collaborative perception has emerged as a promising approach to overcome the limitations of individual perception by enabling multiple agents to exchange information, thus enhancing their situational awareness. Collaborative perception overcomes the limitations of individual sensors, allowing connected agents to perceive environments beyond their line-of-sight and field of view. However, the reliability of collaborative perception heavily depends on the data aggregation strategy and communication bandwidth, which must overcome the challenges posed by limited network resources. To improve the precision of object detection and alleviate limited network resources, we propose an intermediate collaborative perception solution in the form of a graph attention network (GAT). The proposed approach develops an attention-based aggregation strategy to fuse intermediate representations exchanged among multiple connected agents. This approach adaptively highlights important regions in the intermediate feature maps at both the channel and spatial levels, resulting in improved object detection precision. We propose a feature fusion scheme using attention-based architectures and evaluate the results quantitatively in comparison to other state-of-the-art collaborative perception approaches. Our proposed approach is validated using the V2XSim dataset \cite{li2022v2x}. The results of this work demonstrate the efficacy of the proposed approach for intermediate collaborative perception in improving object detection average precision while reducing network resource usage.
\end{abstract}


\section{Introduction}
The recent advancements in deep learning have resulted in remarkable enhancements in the functionality and performance of ITS. These advancements possess the potential to amplify traffic efficiency, minimize traffic accidents, and conserve resources \cite{wen2022deep}. Perception, which is a fundamental module in ITS, enables agents to recognize, interpret, and represent sensory inputs, thus achieving situational awareness. Autonomous driving is one of the extensively researched topics in ITS, which mainly relies on individual perception, where the vehicle senses its surroundings through its own sensors. However, the reliability of such systems depends on the robustness of the onboard sensory and perception units \cite{wen2022deep,zhang2023perception,alaba2023deep}. Despite the significant progress individual perception has witnessed \cite{wen2022deep,bogdoll2022multimodal,alaba2023deep}, developing a reliable perception system still remains a challenge. This is primarily due to issues such as sensor characteristics, obstacle occlusion, illumination, and bad weather conditions. Furthermore, the vehicle’s perception range is generally limited to the range of its sensors, resulting in blind spots in the field of view and making it difficult to provide a full range of perception information for the autonomous vehicle. Consequently, autonomous vehicles fail to detect imminent danger in a timely manner. 

The limitations of single-vehicle perception have led to the emergence of vehicle-to-everything (V2X) communication technology as a promising solution \cite{cui2022cooperative}. V2X communication facilitates communication and data sharing between a vehicle and other entities, including vehicle-to-vehicle (V2V) and vehicle-to-infrastructure (V2I) communication. Collaborative perception techniques have been developed in the domain of V2X to upgrade single-vehicle perception to collaborative perception by incorporating additional viewpoints and fusing data received from multiple connected agents to improve the perception capability of autonomous vehicles. This allows them to see further, better, and even through occlusions \cite{cui2022cooperative,wang2020v2vnet,li2021learning}.

In general, collaborative perception techniques can be classified into three categories based on the type of information exchanged among participants: early, intermediate, and late fusion. Early fusion involves fusing the raw sensor data from each agent into a holistic perspective \cite{chen2019cooper,arnold2020cooperative,ahmed2022joint}. This method provides a complete and detailed view of the environment, but it requires significant communication bandwidth. Late fusion, on the other hand, involves fusing each agent's perception outputs after processing the raw sensor data \cite{miller2020cooperative,allig2019alignment}. This method is bandwidth-efficient but may result in noisy and incomplete fusion results. Intermediate fusion involves extracting features from each sensor modality mounted on each agent and aggregating them into a single representation \cite{chen2019f,wang2020v2vnet,liu2020when2com,liu2020who2com,li2021learning}. This approach is more bandwidth efficient than early fusion and has the potential to achieve both communication bandwidth efficiency and improved perception ability. However, a poorly designed intermediate collaboration strategy may result in information loss during feature abstraction and fusion, leading to limited improvement in perception ability.

This paper proposes a novel approach for designing an effective intermediate collaboration strategy in multi-agent collaborative perception among connected agents. Specifically, we propose a graph attention feature fusion-based neural network, where each node in the graph represents the intermediate representation received from every agent and each edge reflects the pair-wise collaboration between two agents. In addition, we incorporate channel and spatial attention mechanisms to guide the model in identifying relevant features that can enhance the feature map fusion. Our main contribution lies in the development of this new model, which has the potential to significantly improve the perception capability of autonomous vehicles by leveraging collaborative perception. The contribution of this work can be summarized as follows: 
\begin{itemize}
    \item  We have proposed an intermediate collaborative perception method for object detection that utilizes information from neighboring vehicles and infrastructure.
    
    \item We have developed an aggregation methodology that combines intermediate features and uses a spatial and channel attention module to learn relationships between the aggregated features, that enhanced the representation power of the network by guiding it on ``what" and ``where" to focus through the attention module.
    
    \item We have Validated the proposed method using V2X-Sim 2.0  \cite{li2022v2x}, a large-scale V2V and V2X dataset that includes LiDAR data from vehicles and infrastructure, and demonstrated the effectiveness of the proposed method for object detection tasks, showing improved performance compared to existing methods.
\end{itemize}

The remainder of this paper is structured as follows. Section \ref{related_work} presents a summary and analysis of collaborative perception fusion methods. Section \ref{methodology} introduces the proposed methodology, including its primary components and feature fusion strategy. Section \ref{implementation_details} describes the experiment setup, evaluation metric, and baseline models. In Section \ref{results_diss}, the dataset and evaluation metrics are briefly discussed, and the results of our proposed framework are compared to baseline methods. Finally, Section \ref{conclusion} concludes this work.

\section{Related Works}
In this section, we provide a comprehensive review of collaborative perception approaches and systematically summarize the collaboration modules in perception networks.
\\
\newline
\textbf{Early collaborative perception-}
The Early Collaboration approach involves fusing raw sensor data at the input level of the network by projecting all LiDAR point clouds into the ego-vehicles' coordinate frame based on shared pose information among agents. The ego vehicle then aggregates all received point clouds and feeds them to the detector for cooperative perception. Cooper \cite{chen2019cooper} proposes using shared raw point cloud data for cooperative perception and references a neural network for object detection in low-density point cloud data. Another approach, as described in \cite{arnold2020cooperative}, involves a new spatial transformation equation that uses concatenation to fuse sensor data. Building on Cooper's work, \cite{ahmed2022joint} utilizes raw point cloud fusion in which each vehicle compresses the raw point cloud to reduce its size and then shares a cooperative perception message (CPM) that includes the compressed point cloud and location information in the environment. Upon receiving sensor data from neighboring agents, the point clouds are transformed into the receiving vehicle's coordinate system and concatenated with the ego and received data. The resulting aggregated point clouds are then fed into the perception network for processing.
\\
\newline
\textbf{Intermediate collaborative perception -}
The Intermediate Collaboration approach is a fusion technique that combines intermediate features generated by the individual prediction models of each agent, allowing the transmission and subsequent fusion of these features to produce perception outcomes. To this end, Chen et al. \cite{chen2019f} proposed the LIDAR-based F-Cooper method for cooperative object detection, which compares the Voxel Feature Fusion (VFF) and Spatial Feature Fusion (SFF) techniques and examines the impact of sharing a portion of the feature map channels. Although subsequent studies have applied the attention mechanism, a popular deep learning technique, for feature fusion, Who2com \cite{liu2020who2com} utilizes attention to minimize bandwidth usage while selecting a vehicle to communicate with. In contrast, When2com \cite{liu2020when2com} uses attention to determine the communication situation and fuses feature maps from multiple vehicles using a concatenation-based rule. DiscoNet \cite{li2021learning} utilizes matrix-valued attention scores to highlight areas with important information for feature map fusion and employs knowledge distillation (KD) to train the model to resemble the feature maps of the input sharing. V2VNet \cite{wang2020v2vnet} uses a graph neural network (GNN) to represent the topology of multiple vehicles and perform fusion through message passing between neighboring nodes, with two CNNs correcting errors before feature map fusion. Specifically, one CNN compensates for time delay, while the other learns the transformation process to implicitly correct the localization error.
\\
\newline
\begin{figure*}[]
    \centering
    \includegraphics[width=\textwidth]{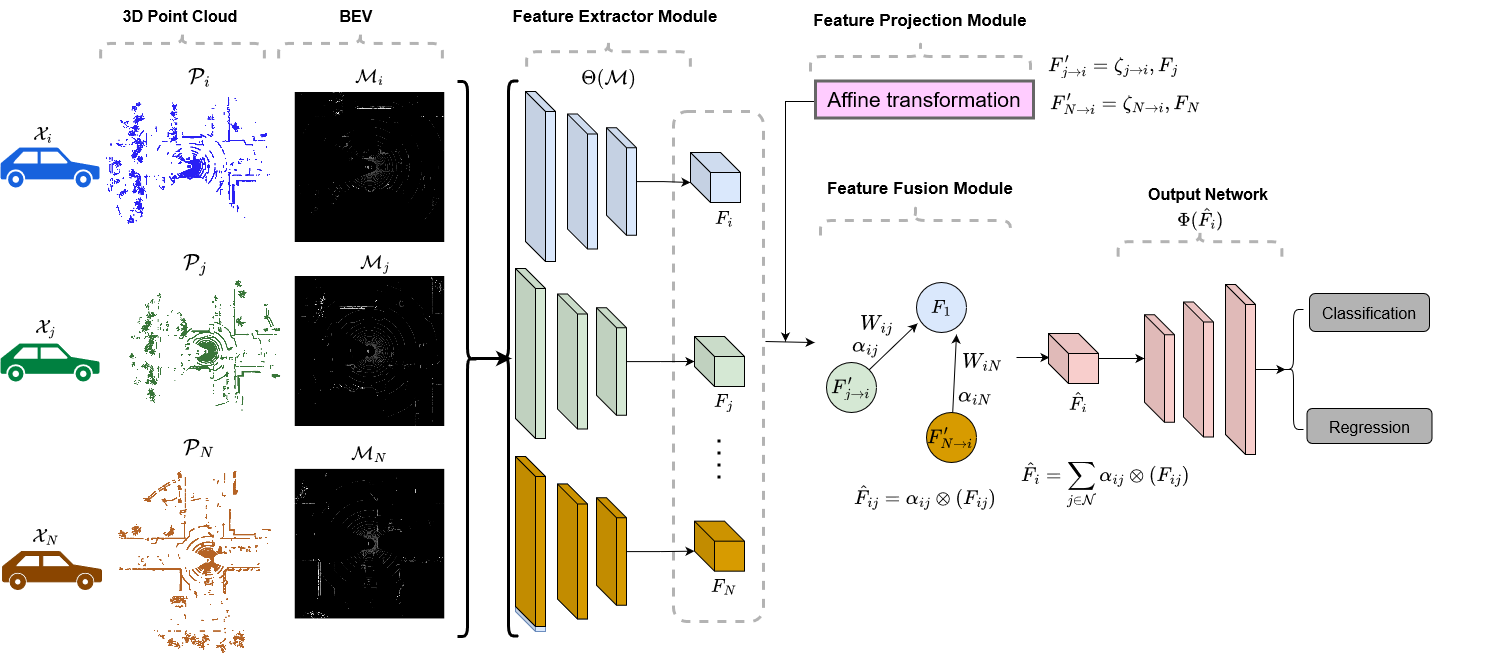}
    \caption{The proposed method's overall architecture begins with each agent $\mathcal{X}$ converting its point cloud $\mathcal{P}$ into a bird's eye view (BEV) map $\mathcal{M}$. The shared feature extractor $\mathcal{E}$ processes the BEV map to obtain the feature map $F$. The received feature maps are then transformed to the ego agent's coordinate system before being aggregated $\mathcal{A}$ with the ego feature map, utilizing channel and spatial attention mechanisms to produce an updated representation $\hat{F}_{i}$. The updated representation is then fed to the output network $\mathcal{D}$ to perform object detection.}
    \label{model_overview}
\end{figure*}
\noindent
\textbf{Late collaborative perception -} 
Late collaboration is a concept that pertains to the merging of perception results in the output space, which is usually carried out after each individual agent has processed its own observations. Miller et al. \cite{miller2020cooperative} utilized this technique to design a perception and localization system for self-driving vehicles, with a focus on addressing issues associated with communication link dropouts and latency. Allig et al. \cite{allig2019alignment} explored the temporal and spatial alignment of shared detected objects and suggested utilizing a non-predicted sender state for transformation, thus eliminating the need for sender motion compensation. Nonetheless, despite the advantage of higher communication bandwidth efficiency of late collaboration, it is vulnerable to positioning errors, which can cause high estimation errors and noise due to incomplete local observations.
\\
\newline
\indent In summary, based on a comparison of three collaboration strategies, intermediate collaboration is found to be more efficient in terms of communication bandwidth usage and results in highly accurate perception outcomes that outperform those of late collaboration strategies \cite{cui2022cooperative}. However, implementing intermediate collaborative perception poses significant algorithmic challenges in practice, such as selecting the most beneficial and concise features for transmission between agents and fusing the received features to improve perception capabilities. Addressing these challenges requires the development of efficient and effective methods for feature selection and fusion that can optimize performance in the collaborative perception process.
 \label{related_work}

\section{Methodology} \label{methodology}
This paper proposes a collaborative fusion architecture for multi-modal feature fusion that utilizes attention mechanisms to selectively combine intermediate representations obtained from multiple agents. The proposed methodology comprises three stages: feature extraction, attention-based feature fusion, and output prediction network. In the feature extraction stage, point cloud data is used to extract relevant features. The resulting feature maps are then passed to the attention-based feature aggregation stage, which selectively fuses the most informative features from each intermediate representation obtained from different agents, leveraging attention mechanisms. Finally, in the output prediction stage, that utilizes the fused features to predict the bounding boxes for object detection.

\subsection{Feature Extraction} \label{Feature Extraction}
3D Point clouds are first captured by the LiDAR sensor deployed on every agent $\mathcal{X} \{i,j..N\}$, where $\{i,j,.., N\}$ are the connected agents. Subsequently, each agent creates a local binary Bird’s-Eye-View (BEV) map $\mathcal{M} \in \mathbb{R}^{W_m \times H_m}$ from the raw point cloud, where $W_{\mathcal{M}}$ and, $H_{\mathcal{M}}$ indicate the width, height of the local BEV map respectively as shown in Fig. \ref{model_overview}.

Subsequently, the BEV of the $i$th agent $\mathcal{M}_{i}$ is passed through the feature extractor which aims to extract informative features BEV provided by the agent obtaining the feature map $F_i \leftarrow \Theta(\mathcal{M}_{i})$, where $\Theta(\cdot)$ is the feature extractor, and  $F_{i} \in \mathbb{R}^{W \times H \times C}$, where $W, H$ and $C$ are the feature dimension. The goal of the feature extractor is to transform the input data and extract features that can be processed while preserving the information in the original data. The proposed method considers homogeneous intermediate collaborative perception to perform object detection, where multiple agents share the same model components, hence, all agents share the same $\Theta(\cdot)$ design. Additionally, the main goal of the proposed method is to enhance the feature map fusion and aggregation strategy and compare our aggregation method with the state-of-the-art models. Therefore we benchmark our aggregation methods utilizing the feature extractor of DicoNet $\Theta_{DiscoNet}(\cdot)$ and V2VNet $\Theta_{V2VNet}(\cdot)$ in order to independently investigate the performance of the aggregation strategy regardless of the feature extractor.

\subsection{Feature projection}
Due to the distinct geological locations and orientations of the agents, the resulting feature maps need to be transformed into the receiver's coordinate system for feature fusion. Affine transformation, which preserves collinearity and ratios of distances, can be employed for this purpose \cite{wakahara2001affine}. Affine transformation refers to a specialized category of projective transformations that does not relocate any objects from the affine space to the plane at infinity or vice versa. To accomplish this, each agent transmits its feature map $F$ alongside pose information $\zeta$, which indicates the location and orientation of the agent, represented as (X, Y, Z, roll, yaw, pitch). Once the $i$th agent receives the $F_j
$ and $\zeta$ from $j$th agent, it projects the $F_j$ into its coordinate system using the transformation $\zeta_{j \rightarrow i}$, which is based on the ego poses $\zeta_i$ and $\zeta_j$ of the receiving and transmitting agents shown in Fig. \ref{model_overview}. The resulting transformed feature map of agent $j$, $F'_{j\rightarrow i}$, is then represented in the same coordinate system as the feature map $F_i$ obtained from the $i$-th agent. Once all the feature maps received from neighboring agents have been transformed into the $i$-th agent's coordinate system, the ego and the transformed feature maps are ready to be processed by the aggregation stage to obtain a final representation that integrates information from all the agents.
\subsection{Feature Fusion}\label{feature_fusion}
\begin{figure}
    \centering
    \includegraphics[width=0.8\columnwidth]{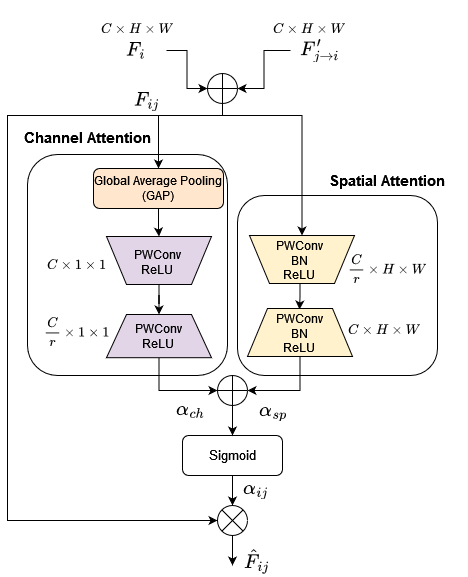}
    \caption{Illustration of the proposed channel-spatial attention feature fusion.}
    \label{attention}
\end{figure}

The previous literature has extensively investigated the significance of attention in feature fusion, as reported in \cite{guo2022attention}. Attention not only directs focus but also enhances the representation of interests. In our proposed feature fusion strategy, inspired by the prominent attention-based feature fusion modules SENet \cite{hu2018squeeze}, MS-CAM \cite{dai2021attentional} and CBAM \cite{woo2018cbam}, we aim to leverage both channel and spatial attention to exploit their respective benefits to increase representation power and fuse intermediate representations from multiple connected agents. The attention mechanism focuses on important features for fusion, suppresses unnecessary ones, and adaptively selects crucial regions. The first step of our feature fusion is to aggregate the feature maps of interest i.e. of the ego $F_i$ and the transformed feature map of the neighboring agent $F'_{j\rightarrow i}$, as shown below:
 \begin{equation}
    F_{ij} = F_i \uplus F'_{j\rightarrow i}
    \label{updated_feature_map_1}
\end{equation}
where $\uplus$ denotes the aggregation process which is either concatenation or addition as shown in Table. \ref{internal_comp} and will be elaborated later in this paper.
\\
\newline
\textbf{Channel Attention}
To generate a channel attention map, we exploit the inter-channel relationship of features, where each channel of a feature map acts as a feature detector \cite{zeiler2014visualizing}. We utilize channel attention as it focuses on determining ``what" is significant within an input-aggregated feature map. Given an intermediate feature $F_{ij}$, to compute the channel attention squeeze the spatial dimension of the aggregated feature map using global average pooling (GAP). Then, the channel attention operation recalibrates the weight of each channel to determine what requires attention. To achieve this we utilize an encoder-decoder point-wise convolution (PwConv) \cite{hua2018pointwise}
to generate the channel attention map $\alpha_{ch}$.
This operation applies $1 \times 1$ convolutional filters to the input feature map, which allows for the adjustment of channel-wise features while keeping the spatial dimensions of the feature map the same. The channel attention weights $\alpha_{ch} \in \mathbb{R}^{C}$ is computed as:
\begin{equation}
    \alpha_{ch} = \delta(\textbf{L}_{\textbf{ch}}(\delta \textbf{G}_{\textbf{ch}}(GAP(F_{ij}))))
    \label{channel attention}
\end{equation}
where $GAP(F_{ij})\in \mathbb{R}^{C}$ denotes the global average pooling of the aggregated feature map $F_{ij}$. $\delta$ denotes the Rectified Linear Unit (ReLU) \cite{nair2010rectified}. The channel attention feature map is computed by an autoencoder with a series of PwConv, where $G_{ch}$ is the PWConv-based encoder representing the dimensionality reduction layer, and $L_{ch}$ is the PWConv-based decoding representing the dimension-increasing layer $r$ is the channel reduction ratio as shown in Fig. \ref{attention}. We can see that the channel attention squeezes each feature map of size $H \times W$ into a scalar $\alpha_{ch} \in R $. 
\\
\newline
\textbf{Spatial Attention}
In addition to channel attention, our feature fusion model also incorporates spatial attention which serves as an adaptive spatial region selection mechanism, guiding the model to ``where" to direct its focus. We generate a spatial attention map by utilizing the inter-spatial relationship of features. We employ autoencoder-based PwConv operation,  as shown in Fig. \ref{attention}, to extract informative features by combining cross-channel and spatial information and generate the spatial attention map $\alpha_{sp}$. Because PwConv only operates on a single spatial location at a time with kernel $1 \times 1$, it can be computed very efficiently, even for very large feature maps. This makes it well-suited for use in spatial attention mechanisms, where attention weights need to be computed for each spatial location in the feature map. This enables the aggregation module to emphasize significant features along these two main dimensions: channel and spatial axes.

On the aggregated feature maps $F_{ij}$, we apply an autoencoder-based PWConv as shown in Fig. \ref{attention}, to generate a spatial attention map $\alpha_{sp} \in \mathbb{R}^{C \times H \times W}$ which is computed as follows:
\begin{equation}
    \alpha_{sp} =\beta(\delta(\textbf{L}_{\textbf{sp}}(\beta(\delta(\textbf{G}_{\textbf{sp}}(F_{ij}))))))
    \label{spatial attention}
\end{equation}
where $\beta$ denotes the Batch Normalization (BN) \cite{ioffe2015batch}, and $\uplus$ represents concatenation or summation operation. 

We incorporate channel and spatial attention modules separately, as illustrated in Fig. \ref{attention}, so that each branch can learn ``what" and ``where" to focus on in the channel and spatial axes, subsequently, the attention maps are combined to leverage those learnings. Subsequently, the computed channel and spatial attention feature maps $\alpha_{ch}$ and $\alpha_{sp}$, respectively, are summed and passed through sigmoid function $\sigma$  to create $\alpha(F_{ij}) \in \mathbb{R}^{C\times H\times W}$ which denotes the total attention weights:
\begin{subequations}
\begin{equation}
    \alpha_{ij} = \sigma(\alpha_{ch} \oplus \alpha_{sp})
\end{equation}
\begin{equation}
   \hat{F}_{ij}= \alpha_{ij} \otimes (F_{ij})
\end{equation}
\end{subequations}
This process is repeated for every feature map received, and the updated fused feature map  of the $i$th agent $\hat{F}_{i} \in \mathbb{R}^{C \times H \times W}$ is computed as:
\begin{equation}
     \hat{F}_i = \sum_{j\in \mathcal{N}}\alpha_{ij}\otimes(F_{ij})
\end{equation}
%%VIP
%%https://reader.elsevier.com/reader/sd/pii/S0950705122000582?token=AC3CB6AD15EAC4CA0D1F87C00B0EFB94BC069AE0ADF6C72DA151560C8742DC82235A5133BD40731E66C8444578F089A5&originRegion=eu-west-1&originCreation=20230221101208

%https://reader.elsevier.com/reader/sd/pii/S0168169922003660?token=B2BEA2A85512C2CDD0F48CAF6307416BC1FFBC840B20F7DF70F7596708FB1B46BE710C4F89DCECFFC49D6D4D3E838AA4&originRegion=eu-west-1&originCreation=20230221101206
\subsection{Output Network} \label{output_netowrk}
After collaboration and feature map aggregation, each agent passes its updated feature map to the output network to generate the final detection outputs. The decoded feature map is  ${\hat{y}, \hat{x}}\leftarrow {\Phi}(\hat{F}_i)$, the $\Phi(.)$ is composed of convolution layers to classify the foreground-background categories and regress the bounding boxes. Following the reasoning discussed in section \ref{Feature Extraction} we utilize the same output network used by DiscoNet $\Phi_{DiscoNet}$ \cite{li2021learning} and V2VNet $\Phi_{V2VNet}$ \cite{wang2020v2vnet} to generate the final detection outputs.

\section{Implementation Details} \label{implementation_details}
\noindent
\textbf{Dataset}
V2X-Sim 2.0 \cite{li2022v2x} is a vehicle-to-everything simulation-based dataset that has been constructed by integrating SUMO \cite{krajzewicz2012recent} and Carla \cite{Dosovitskiy17}. The simulation leverages SUMO to generate realistic traffic flow data, and Carla is utilized to obtain sensory streams, such as LiDAR data, from multiple vehicles that are located within the same geographic area. The dataset is composed of 10,000 frames that are distributed across 100 distinct scenes, with each scene consisting of 100 frames. The dataset has been partitioned with a split of 8,000/1,000/1,000 frames being designated for the training, validation, and testing sets, respectively. Each frame is comprised of data collected from several agents, including vehicles and roadside units (RSUs). The dataset encompasses 37,200 training samples, 5,000 validation samples, and 5,000 test samples. Our experimental evaluation is conducted under two scenarios: without (w/o) and with (w/) RSU as shown in Table. \ref{ext_comp}. For this work, we only target vehicle detection and report the results on the test set. 

\\
\newline
\textbf{Evaluation}
We define the total model loss function as a weighted sum of the classification loss $L_{cls}$  and the regression loss $L_{reg}$ as follows:
\begin{equation} 
    L_{total} = \rho L_{cls} + \psi L_{reg}
      \label{total_loss}
\end{equation}
where $\rho$ and $\psi$ are hyperparameters defining the weights of the classification loss and regression loss.
For the classification loss $L_{cls}$ we employ binary cross-entropy loss to supervise foreground-background classification:
\begin{equation}
    L_{cls}(y,\hat{y}) = -(ylog(\hat{y}) + (1-y)log(1-\hat{y}))
    \label{focal_loss}
\end{equation}
where $y$ is the true label (0 or 1) of the input, and $\hat{y}$ is the predicted probability of the input belonging to class 1. For the regression loss $L_{reg}$, the $smooth_{L_1}$ loss is used to supervise the bounding-box regression: 
\begin{equation}
    smooth_{L_1}(x) = 
    \begin{cases}
    0.5(\hat{x} - x)^2 & \text{if $\lvert \hat{x} - x \rvert$ $ < 1$ } \\
    \lvert \hat{x} - x \rvert - 0.5 & \text{otherwise,}
    \end{cases}  
    \label{smooth_l1}
\end{equation}
where $x$ is the difference between the predicted and ground-truth values for a single coordinate of the bounding box. To compute the smooth L1 loss for the entire bounding box, the loss is typically computed separately for each coordinate (e.g., the x-coordinate, y-coordinate, width, and height) and then summed together. The smooth L1 loss is a good choice for bounding box regression because it is less sensitive to outliers than other loss functions like the L2 loss, which leads to more stable and accurate predictions.
% \usepackage{multirow}
\begin{table}[ht]
\caption{Architectures for our proposed method. Where ``Aggregation Operation" represents the $\uplus$ featured in Eq. \ref{updated_feature_map_1}. ``Layer" represents the feature map dimensionality reduction (see also Fig. \ref{attention}, Eq. \ref{channel attention} and \ref{spatial attention})}.
\centering
\begin{tabular}{clcl}
\hline
\multirow{2}{*}{\textbf{Model Base}} & \multicolumn{1}{c}{\multirow{2}{*}{\textbf{\begin{tabular}[c]{@{}c@{}}Experiment \\ Name\end{tabular}}}} & \multirow{2}{*}{\textbf{\begin{tabular}[c]{@{}c@{}}Aggregation\\ Operation ($\uplus$)\end{tabular}}} & \multicolumn{1}{c}{\multirow{2}{*}{\textbf{Layers}}} \\
 & \multicolumn{1}{c}{} &  & \multicolumn{1}{c}{} \\ \hline
\multirow{4}{*}{\textbf{DiscoNet}} & $FF_{D}$\_Sum64 & Sum & 256, 128, 64 \\
 & $FF_{D}$\_Sum32 & Sum & 256, 128, 64, 32 \\
 & $FF_{D}$\_Concat64 & Concat & 512, 256, 128, 64 \\
 & $FF_{D}$\_Concat32 & Concat & 512, 256, 128, 64, 32 \\ \hline
\multirow{2}{*}{\textbf{V2VNet}} & $FF_{V}$\_Concat64 & Concat & 512, 256, 128, 64 \\
 & $FF_{V}$\_Concat32 & Concat & 512, 256, 128, 64, 32 \\ \hline
\end{tabular}
\label{internal_comp}
\end{table}
% \usepackage{multirow}
\begin{table*}[]
\caption{Quantitative results of BEV detection on V2X-SIM 2.0. AP denotes average perception with and without RSU at IoU of 0.5 and 0.7. $\Delta_{size}$ is the delta size $\%$ increase $\uparrow$ and reduction $\downarrow$ from the model when compared to the state-of-the-art intermediate collaboration V2VNet and DiscoNet.}
\centering
\begin{tabular}{lccccccl}
\hline
\multirow{2}{*}{Method} & \multicolumn{2}{c}{AP@IOU = 0.5} & \multicolumn{2}{c}{AP@IOU = 0.7} & \multirow{2}{*}{\textbf{\begin{tabular}[c]{@{}c@{}}Model Size\\ (Mb)\end{tabular}}} & \multirow{2}{*}{\textbf{\begin{tabular}[c]{@{}c@{}}$\Delta_{size}$ \% \\ V2VNet\\ \end{tabular}}} & \multicolumn{1}{c}{\multirow{2}{*}{\textbf{\begin{tabular}[c]{@{}c@{}}$\Delta_{size}$ \% \\ DiscoNet\\ \end{tabular}}}} \\ \cline{2-5}
 & \textbf{w/o RSU} & \textbf{w/ RSU} & \textbf{w/o RSU} & \textbf{w/ RSU} &  &  & \multicolumn{1}{c}{} \\ \hline
Lower-bound\cite{li2022v2x} & 49.9 & 46.96 & 44.21 & 42.33 & 90.5 & $\downarrow$ 50.11 &  \multicolumn{1}{c}{$\downarrow$ 25.45} \\
Co-lower-bound\cite{li2022v2x} & 43.99 & 42.98 & 39.1 & 38.26 & 90.5 & $\downarrow$ 50.11 & \multicolumn{1}{c}{$\downarrow$ 25.45} \\
Upper-bound\cite{li2022v2x} & 70.43 & 77.08 & 67.04 & 72.57 & 90.5 & $\downarrow$ 50.11 & \multicolumn{1}{c}{$\downarrow$ 25.45} \\
When2com\cite{liu2020when2com} & 44.02 & 46.39 & 39.89 & 40.32 & 247.5 & $\uparrow$  51.32 &  \multicolumn{1}{c}{$\uparrow$ 126.11} \\
When2com*\cite{liu2020when2com} & 45.35 & 46.39 & 40.45 & 41.43 & 247.5 & $\uparrow$  51.32 &  \multicolumn{1}{c}{$\uparrow$ 126.11} \\
Who2com\cite{liu2020who2com} & 44.02 & 46.39 & 39.89 & 40.32 & - & - & \multicolumn{1}{c}{-} \\
Who2com*\cite{liu2020who2com} & 45.35 & 48.28 & 40.45 & 41.13 & - & - & \multicolumn{1}{c}{-} \\
V2VNet\cite{wang2020v2vnet} & 68.35 & 72.08 & 62.83 & 65.85 & 181.4 & - & \multicolumn{1}{c}{$\uparrow$ 49.42} \\
DiscoNet\cite{li2021learning} & \textbf{69.03} & 72.87 & 63.44 & 66.4 & \textbf{121.4} & $\downarrow$ 33.07 & \multicolumn{1}{c}{-}\\
\hline
$FF_{D}$\_Sum64 (Ours) & 68.14 & 71.81 & 62.36 & 65.86 & 122.5 & $\downarrow$ 32.47 & \multicolumn{1}{c}{$\uparrow$ 0.91} \\
$FF_{D}$\_Sum32 (Ours) & 68.56 & \textbf{72.96} & \textbf{63.48} & \textbf{65.94} & 122.7 & $\downarrow$ 32.36 & \multicolumn{1}{c}{$\uparrow$ 1.07} \\
$FF_{D}$\_Concat64 (Ours) & 68.97 & 71.61 & 62.88 & 64.41 & 122.7 & $\downarrow$ 30.76 & \multicolumn{1}{c}{ $\uparrow$ 3.46} \\
$FF_{D}$\_Concat32 (Ours) & 68.50 & 72.25 & 63.32 & 63.74 & 126 & $\downarrow$ 30.54 & \multicolumn{1}{c}{ $\uparrow$ 3.79} \\
$FF_{V}$\_Concat64 (Ours) & 67.53 & 70 & 61.55 & 63.52 & 126.6 & $\downarrow$ 30.20 & \multicolumn{1}{c}{$\uparrow$ 4.28} \\
$FF_{V}$\_Concat32 (Ours) & 68.46 & 70.94 & 63.10 & 63.15 & 126.6 & $\downarrow$ 30.70 & \multicolumn{1}{c}{$\uparrow$ 3.52} \\ \hline
\end{tabular}
\label{ext_comp}
\end{table*}
\\
\newline
\textbf{Baseline models}
Our objective is to evaluate the efficacy of our collaborative perception strategy against established baseline collaboration perception models, which are categorized as no, early, intermediate, and late collaboration. For the no collaboration model, we use the Lower-bound model \cite{li2022v2x}, which is a single-agent perception model that processes a single-view point cloud without collaboration. For the early collaboration model, we use the Upper-bound model \cite{li2022v2x}, which fuses raw point cloud data from all connected agents. Intermediate models, including DiscoNet \cite{li2021learning}, V2VNet \cite{wang2020v2vnet}, F-Cooper \cite{chen2019f}, When2com \cite{liu2020when2com}, and Who2com \cite{liu2020who2com}, are based on the communication of intermediate features of the neural network. For the late collaboration model, we consider the Co-lower-bound model \cite{li2022v2x}, which fuses the output predictions from different single-agent perception models. For comparison with the baseline models, we utilize the Intersection-over-Union (IoU) thresholds of  0.5 and 0.7 are selected for calculating average precision thresholds.
\\
\newline
\textbf{Implementation Details} 
As mentioned in sections. \ref{Feature Extraction} and \ref{output_netowrk}, we adopt the same feature extractor as well as the output network as DicoNet \cite{li2021learning} and V2VNet \cite{wang2020v2vnet} as discussed earlier in section. \ref{Feature Extraction}. Moreover, all of the methods are trained with the same setting to ensure that the performance gain is the result of our proposed collaboration strategy rather than other modifications in the model or the architecture. Additionally, we conduct collaboration at the same intermediate feature layer. We utilize the Adam optimizer \cite{kingma2014adam} with an initial learning rate of $10^{\textminus3}$ and steadily decay at every 10 epochs using a factor of 0.1. All models are trained on Tesla V100.

\section{Results and Discussion} \label{results_diss}
Tables. \ref{ext_comp} and \ref{internal_comp} shows the comparison in terms of AP(@ IoU 0.5 and 0.7) for internal methods and baseline methods. As presented in Section \ref{feature_fusion}, a convolution-based encoder-decoder attention mechanism was employed to compute the attention weights to identify significant regions in the received feature map that can enhance the ego agent's perception. In this regard, Table. \ref{internal_comp} presents two encoding-decoding layer depths that were utilized, namely (256, 128, 64) and (256, 128, 64, 32), to investigate the effect of a deeper network on attention weights learned, overall model performance, and size. Notably, the $\uplus$ notation, (presented in Eq. \ref{updated_feature_map_1}) is used to denote the aggregation operation representing the concatenation or addition operation as shown in Table. \ref{internal_comp}, is used to examine and compare the effect of those operations on the initial aggregation stage. The results of our experiments, presented in Table \ref{ext_comp}, indicate that the $FF_D\_Sum64$ model achieves the highest detection precision while maintaining a low model parameter count, which aligns with the hypothesis proposed by \cite{guo2022attention} that feature addition yields better fusion results. Furthermore, it can be observed that a deeper encoder-decoder leads to higher AP due to the fact that deeper layers in the network have access to higher-level, more abstract representations of the input, which can allow the network to learn more complex patterns and relationships between features. This is because the deeper layers are able to combine features learned in earlier layers to form higher-level representations that capture more complex aspects of the input data which is very vital to learn the attention weights. However it was also found that doing deeper (beyond the tested depth presented in Table. \ref{internal_comp}) resulted in lower AP due to the vanishing gradient problem, which occurs when the gradient signal is too small to propagate through multiple layers. This makes it challenging to update the weights of early layers in the network, leading to poor convergence.
 
It is evident from the results that collaboration significantly improves the performance over no collaboration (lower bound) in all cases. Among the collaboration methods, the upper-bound (early collaboration) outperforms all other methods, although it poses challenges with the shared data size which is around 860kb per sweep when compared to the co-lowerbound, intermediate collaboration which is around 5kb and 260kb per sweep, respectively. The intermediate collaboration methods exhibit superior performance compared to the co-lower bound. In terms of terrain and road categories, the proposed methods in this paper (Table. \ref{internal_comp}), V2VNet, DiscoNet, and upper-bound achieve similar performance. V2V and V2I (w/RSU) together generally enhance the perception of V2V only with more viewpoints, which is evident throughout the intermediate models presented in Table .\ref{ext_comp}, and that's due to the availability of more data to train the model on.

The proposed attention-based collaboration, which utilizes well-designed channel and spatial attention-based fusion, achieves highly competitive performance among the intermediate models. The results are listed in Table. \ref{ext_comp} indicate that the proposed collaboration can often surpass the performance of V2VNet, such as in the cases of $FF_{D}\_Sum64$, $FF_{D}\_Sum32$, $FF_{D}\_Concat64$, and $FF_{D}\_Concat32$. Nevertheless, our proposed methods consistently result in much smaller parameter budgets, as evidenced by the ``$\delta_{size} \% (V2VNet)$'' column, which shows a maximum and minimum delta of 33.07\% and 30.20\%, respectively. Compared to DiscoNet, our proposed methods utilizing the DiscoNet base $FF_D$ achieve similar or slightly better performance, such as in the case of $FF_D\_Sum64$.

\section{Conclusion}\label{conclusion}
In this paper, a channel-spatial attention-based approach is proposed for achieving intermediate fusion in collaborative perception. The effectiveness of the proposed approach is demonstrated by benchmarking several state-of-the-art collaborative perception methods in collaboration. Additionally, the results are validated using V2X-Sim 2.0, a large-scale multi-agent 3D object detection dataset. The proposed methods achieve comparable results with the state-of-the-art and, in some cases, surpass them in terms of the number of parameters in the model or average precision.

\section*{Acknowledgement}
This work was supported by the Research Foundation Flanders (FWO) under Grant Number 1S90022N. This work will be validated using 5G network within the Horizon 2020 5G-Blueprint project, funded by the European Commission under agreement No. 952189.

\bibliographystyle{unsrt}
\bibliography{References}
\end{document}